\documentclass[twocolumn,times,tighten,twocolappendix,resetfootnote]{aastex7}
\usepackage{graphicx,amssymb,amsmath,amsthm,amsfonts,epsfig}
\hypersetup{colorlinks=true, citecolor=blue, linkcolor=red, urlcolor=red}

\newcommand{\Po}{P_{\text{orb}}}
\newcommand{\fGW}{f_{\rm GW}}
\newcommand{\Mc}{M_{\rm c}}
\newcommand{\Ms}{M_{\star}}

\definecolor{red}{rgb}{0.71, 0.25, 0.15}

\begin{document}

\title{Revealing the nature of long-period transients with space-based gravitational-wave interferometers}

\author{Arthur G. Suvorov}
\email[show]{arthur.suvorov@tat.uni-tuebingen.de}
\affil{Departament de F{\'i}sica Aplicada, Universitat d'Alacant, Ap. Correus 99, E-03080 Alacant, Spain}
\affil{Theoretical Astrophysics, IAAT, University of T{\"u}bingen, T{\"u}bingen, D-72076, Germany}

\author{Clara Dehman}
\email{clara.dehman@ua.es}
\affil{Departament de F{\'i}sica Aplicada, Universitat d'Alacant, Ap. Correus 99, E-03080 Alacant, Spain}

\author{Jos{\'e} A. Pons}
\email{jose.pons@ua.es}
\affil{Departament de F{\'i}sica Aplicada, Universitat d'Alacant, Ap. Correus 99, E-03080 Alacant, Spain}

\begin{abstract}
A few members of the recently discovered class of long-period transients have been identified as binaries with white-dwarf primaries. In most cases, however, electromagnetic data are inconclusive, and isolated magnetars or compact binaries remain viable. If the pulsation period matches that of the orbit---as is the case for ILT J1101+5521 and GLEAM-X J0704--37---some of these elusive radio transients could be gravitational-wave bright in the mHz band. Space-based interferometers could thus be used to provide independent constraints on their nature. We quantify the signal-to-noise ratio for the known systems under various scenarios and show that a few could be detectable for sufficiently large chirp masses. Astrophysical implications for (non)detections are discussed.
\end{abstract}


\section{Introduction} \label{sec:intro}
\label{firstpage}

{Radio surveys are revealing a sizeable population of what are called `long period transients' (LPTs). 
These systems pulse coherently in the radio band with periods that range from several minutes to many hours and 
exhibit diverse observational properties}.
LPT brightness variations imply compact sources and, given spectral similarities to pulsar emissions, the natural interpretation is that they are {slowly rotating} neutron stars \cite[see][for a discussion]{cz24}. 
However, this is theoretically challenging to accept as their {reported} radio luminosities ($L_{\nu}$) {typically} exceed the implied spindown luminosity \cite[e.g.][]{hw22,men25}, necessitating an alternative power reservoir. 
While magnetic fields could provide such a battery for magnetar progenitors \citep{ben20,coop24}, optical spectroscopy has confirmed that some LPTs are binaries consisting of white dwarfs with M-dwarf companions 
\cite[WDMDs;][]{hw24,rui24,rod25}. 
As LPTs reside within {largely unexplored parameter spaces}, tools to constrain their nature are of astrophysical relevance.

Because pulsations from ILT J1101+5521 and GLEAM-X J0704--37 -- confirmed WDMDs -- are phase-aligned with their orbital motion \citep{rui24,rod25}, 
{this may also apply to other LPTs.}
{However,} the lowest-period systems cannot be WDMDs with {pulsation periods locked to the orbit ($P \approx \Po$), as the companion would not fit within} its Roche lobe. 
{Nevertheless, since} the maser-like radio emission mechanisms often invoked to explain pulsations in binaries do not rely on companion makeup \citep{mel17,qu25}, {LPTs could also involve} a pair of compact objects \citep{dong25,bloot25,lyman25}.

We point out that because several LPTs reside at sub-kpc distances, they may be strong sources of gravitational waves (GWs) independently of the radio mechanism \emph{if} the orbital period corresponds to the pulsation period. As eccentricity should be erased before a binary compactifies to $\Po \sim$~hours \citep{peters64}, emissions at frequencies of $\fGW = 2 /\Po \approx 0.6 \, \text{mHz} \times \left({1 \, \text{hr}}/{\Po}\right)$ ought to dominate the spectrum in this scenario. Within the next decade, a network of space-based, GW interferometers will become operational, namely the Laser Interferometer Space Antenna \cite[LISA;][]{lisa17} plus the two Chinese-led projects Taiji \citep{luo20} and TianQin \citep{huang20}. {We focus on LISA and Taiji in this work, as they} will be most sensitive in the $\gtrsim$~mHz band and {are best-suited for 
detecting}
signals from 
{known} LPTs. {We show that} future GW observations can {be leveraged} to effectively \emph{confirm} or \emph{rule out} binary scenarios in 
{a number of} systems
where electromagnetic data 
{remain inconclusive.}

This paper is organized as follows. Multiwavelength observations and binary scenarios 
are reviewed in Section~\ref{sec:multi}, to pave the way for estimates of the characteristic strains from LPTs in Section~\ref{sec:orbgws}. Signal-to-noise ratios (S/Ns) are quantified in Section~\ref{sec:lisa} {for LISA alone and for a joint network in Section~\ref{sec:network}},
{where it is shown that several LPTs are promising GW detection candidates under favorable conditions. An outlook for future surveys is provided in Section~\ref{sec:future}.} 
Astrophysical implications are also discussed, with conclusions presented in Section~\ref{sec:discussion}.

\section{Constraints on binary nature} \label{sec:multi}
{In this work, we pose the question: \emph{if} LPTs (i) consist of a binary system involving at least one compact object and (ii) pulsate at the orbital period, would they be detectable by space-based GW interferometers?}
{Given that the period is known in these systems, 
the main additional parameters needed for this assessment}
are the distance, $d$, and the chirp mass,
\begin{equation} \label{eq:chirp}
\mathcal{M} = \frac {\left( \Ms \Mc \right)^{3/5}} { \left(\Ms + \Mc \right)^{1/5}} ~,
\end{equation}
for primary and companion masses $\Ms$ and $\Mc$, respectively. While LPT distances are constrained {to within a factor $\sim 2$ by} their dispersion measures \cite[DMs; cf.][]{price21}, system masses (or indeed binary nature) are much more uncertain in most cases. In order to estimate the viable chirp-mass range and the viability of the hypothesis generally, we begin by reviewing the main constraints in this context.

\subsection{White-dwarf plus M-dwarf} \label{sec:wdmd}

Two LPTs---GLEAM-X J0740--34 and ILT J1101+5521---have been confirmed as WDMDs with $P = P_{\rm orb}$ from broadband photometry, optical spectroscopy, and other channels \citep{hw24,rui24,rod25}. Their data respectively indicate $\Mc \approx 0.19 M_{\odot}$ and $\Mc \approx 0.32 M_{\odot}$ \cite[though cf.][{for a Markov chain Monte Carlo parameter exploration}]{rod25}. While primary masses cannot be dynamically-inferred with certainty due to the unknown inclination, the discovery papers suggest $\Ms \sim 0.6 M_{\odot}$ and $\Ms \sim 0.8 M_{\odot}$, respectively. 

For ILT J1101+5521 however, \cite{rui24} note that arguments based on Roche-lobe overflow indicate the tighter bound\footnote{This restriction applies for an M-dwarf radius of $R_{\rm MD} = 0.217 R_{\odot}$. If the radius was lower by even a few percent ($R_{\rm WD} \lesssim 0.21 R_{\odot}$), the primary mass becomes practically unconstrained (see their extended data figure 8).} $\Ms \lesssim 0.3 M_{\odot}$. In particular, observations of M dwarfs indicate a mass-radius relationship of $M_{\rm MD}/M_{\odot} \approx R_{\rm MD} / R_{\odot}$ \citep{pars18}, with the lightest (confirmed) object being EBLM J0555--57AB with $M_{\rm MD} = 0.084(4) M_{\odot}$ \citep{boet19}. Using the \cite{egg} formula and assuming a Keplerian orbit, the radius of the Roche sphere reads
\begin{equation} \label{eq:eggformula}
\begin{aligned}
R_{\rm R} &\approx a\times\frac{0.49 q^{2/3}}{0.6q^{2/3} + \log(1 + q^{1/3})} \\
&\gtrsim 0.1 R_{\odot} \left(\frac{P_{\rm orb}}{60\text{ min}}\right)^{2/3},
\end{aligned}
\end{equation}
for orbital separation $a$ and mass-ratio $q$. Since $R_{\rm MD} \leq R_{\rm R}$ is required for stability, an effective upper limit to $\Ms$ via $q$ is set by expression \eqref{eq:eggformula} if $P = P_{\rm orb}$. 
Such an argument further implies that 
{LPTs with $P \lesssim 30$~min are unlikely to be WDMDs pulsing at the orbital period} (see Table~\ref{tab:orbdata} for our considered sample\footnote{{Note that no distance estimate is provided by \cite{ak25} for ASKAP J1448. They consider a plausible lower limit of $\sim 200$~pc, adopted here, with a maximum of $\sim 10$~kpc.}}). 
{We exclude a WDMD scenario for J0630, GX J1627, and GPM J1839 for this reason (see also 
Sec.~\ref{sec:j1634}). 
For GX J1627 in particular, dedicated spectrophotometric follow-up practically rules out an M dwarf \citep{lyman25}.} 
Alternative scenarios with either more compact progenitors, $P \neq P_{\rm orb}$, or isolated stars are required in these cases.

\begin{table}
\centering
\caption{Pulsation periods (in descending order) and dispersion-measure distances for the LPTs considered in this work. {Full source names, and abbreviations used throughout, are provided in the notes below. For ILT/CHIME J1634+44, two orbital periods have been considered in the literature owing to interpulse/offset phenomena (see Sec.~\ref{sec:j1634}).}}
\hspace{-1.2cm}\begin{tabular}{lcc}
 \hline
 \hline
Source name & Period (s) & Distance (kpc) \\
\hline
CHIME J0630+25$^{\text{a}}$ & 421.4 & 0.17(8)  \\
ILT/CHIME J1634+44$^{\text{b}}$ & 841.2 / 2103.1 & 2.2(8)  \\
GLEAM-X J1627$^{\text{c}}$ & 1091.2 & 1.3(5) \\
GPM J1839--10$^{\text{d}}$ & 1318.2 & 5.7(2.9) \\
DA J1832$^{\text{e}}$ & 2656.2 & 4.8(8) \\
ASKAP J1935$^{\text{f}}$ & 3225.3 & 4.9(5) \\
ASKAP J1755$^{\text{g}}$ & 4186.3 & 4.7(1.3) \\
GCRT J1745--3009$^{\text{h}}$ & 4620.7 & $\sim$ 7(1) \\
ASKAP J1448$^{\text{i}}$ & 5631.1 & $\sim$ 5(4.8) \\
ILT J1101+5521$^{\text{j}}$ & 7531.2 & 0.50(1.3) \\
GLEAM-X J0704--34$^{\text{k}}$ & 10496.6 & 0.4(1) \\
\hline
\hline
\end{tabular}
\\
\noindent \footnotesize{\textbf{Notes.} $\vphantom{\text{a}}^{\text{a}}$CHIME J0630+25 \cite[J0630;][]{dong24} $\vphantom{\text{b}}^{\text{b}}$ILT/CHIME J1634+44 \cite[J1634;][]{bloot25,dong25}  $\vphantom{\text{c}}^{\text{c}}$GLEAM-X J162759.5-523504.3 \cite[GX J1627;][]{hw22} $\vphantom{\text{d}}^{\text{d}}$GPM J1839--10 \cite[GPM J1839;][]{hw23} $\vphantom{\text{e}}^{\text{e}}$DART/ASKAP J1832–0911 \cite[DA J1832;][]{wang24} $\vphantom{\text{f}}^{\text{f}}$ASKAP J1935+2148 \cite[A J1935;][]{caleb24} $\vphantom{\text{g}}^{\text{g}}$ASKAP J175534.9--252749.1 \cite[A J1755;][]{sween25} $\vphantom{\text{h}}^{\text{h}}$GCRT J1745--3009 \cite[GC J1745;][]{hyman09} $\vphantom{\text{i}}^{\text{i}}$ASKAP J144834--685644 \cite[A J1448;][]{ak25} $\vphantom{\text{j}}^{\text{j}}$ILT J1101+5521 \cite[ILT J1101;][]{rui24} $\vphantom{\text{k}}^{\text{k}}$GLEAM-X J0704--34 \cite[GX J0740;][]{hw24}.}
\label{tab:orbdata}
\end{table}

With respect to the possibility that $P \neq P_{\rm orb}$, there are two known white-dwarf pulsars (J1912 and Ar Sco) with periods that \emph{do not} correspond to the orbit but rather the spin of the primary \citep{marsh16,peli23,peli24}. 
However, the radio profiles of these objects are notably different from the two confirmed WDMD LPTs. 
Their observed (highest peak) radio luminosities, $\sim 2 \times 10^{26}$~erg/s at $\approx 1$~GHz, are much lower than the values of ILT J1101 ($\sim 1.5 \times 10^{28}$~erg/s at $\approx$1~GHz) and GX J0740 ($\sim 10^{28}$~erg/s at $\approx$100~MHz); see appendix B.3 of \cite{rod25}. 
Additionally, the degree of linear polarization in J1912 and Ar Sco are at the level of a few percent while the LPTs ILT J1101 and GX J0740 display values of at least $\approx 20\%$ \cite[see table 1 in][]{qu25}. 
Ar Sco and J1912 also shine persistently in X-rays, while LPTs do not {\cite[with the exception of A J1448;][]{ak25}}. 
The operative pulsation mechanisms and binary characteristics 
{are therefore likely different between these astronomical classes; for this reason, we exclude J1912 and AR Sco from our sample {\cite[though the latter may be marginally visible for a multi-year folding-time given its distance of $\sim 110$~pc, chirp mass $\sim 0.5 M_{\odot}$, and orbital period $\approx 3.57$~hours;][]{marsh16}.}}
{We also exclude ASKAP J183950.5--075635.0 (A J1839) from our sample as this source displays prominent interpulses at (main-pulse) phases of $\approx 0.5$ \citep{lee25}. 
While this cannot conclusively rule-out a binary origin for the source \cite[cf.][]{bloot25}, these extra, half-period pulses are more easily explained by an isolated object emitting from both magnetic poles \cite[see][for a magnetar model]{sdp25}.}

For WDMD scenarios, we take a modest value of 
\begin{equation} \label{eq:chirpwdmd}
\mathcal{M}_{\rm WDMD} \approx 0.3 M_{\odot},
\end{equation}
consistent with the mid-range obtained from the orbital solutions for ILT~J1101 ($\mathcal{M} \approx 0.29 M_{\odot}$) and GX~J0704 ($\mathcal{M} \approx 0.43 M_{\odot}$) quoted earlier.

\subsection{White dwarfs with compact companions} \label{sec:wdwd}

If the shortest period LPTs are not WDMDs, what type of systems could 
they be?
While isolated neutron stars are viable -- and perhaps even favored in most cases \citep{ben20,coop24} -- double degenerates are harder to rule out owing to their intrinsic dimness \cite[see][for a discussion]{lyman25}.

In contrast to WDMDs, there are double-compact systems with observed orbital periods shorter than LPTs (e.g. AM CVn stars). For instance, HM Cancri boasts an orbital period of $\approx 322$~s and was inferred to be 
{massive} by \cite{roel10} based on 
{modulation of the helium emission lines, indicating}
a significant chirp mass (see below). 
Moreover,
{it has been proposed that a unipolar inductor may drive the evolution of (at least some) AM CVns \citep{wu02,dall07}, potentially also triggering radio activity\footnote{Although subsequent surveys have not recovered such features, \cite{ram07} found evidence at 5.8$\sigma$ for coherent (brightness temperature $\gtrsim 10^{18}$~K) radio activity in HM Cancri which would be difficult to explain without a unipolar inductor. {For a discussion on radio activity in other cataclysmic variables, see \cite{rid23}.}}}
{if the magnetic primary spins fast enough to create a centrifugal barrier that prevents stray plasma from screening the interaction zone \citep{ill75}.}
\cite{mac24} identified two magnetic dwarfs within AM CVns, supporting the possibility that magnetospheric interactions in similar systems could instigate radio activity \citep{willes04,rid23}.

In scenarios of a double dwarf (or dwarfs with a compact companion which we do not distinguish notationally), we consider a (optimistic) chirp mass of
\begin{equation} \label{eq:chirpwdwd}
\mathcal{M}_{\rm WDWD} \approx 0.5 M_{\odot}.
\end{equation}
The value \eqref{eq:chirpwdwd} is slightly higher than the range inferred for the tight AM CVns HM Cancri ($\mathcal{M} \approx 0.45 M_{\odot}$), V803 Cen ($\mathcal{M} \lesssim 0.26 M_{\odot}$), and V407 Vul \cite[$\mathcal{M} \lesssim 0.35 M_{\odot}$; see table 2 in][]{sol10}, though this may be justified by noting that dwarf masses are correlated with magnetic field strengths \citep{kawla20} and strong fields may be required to instigate radio pulsing. For a more conservative estimate, the reader can instead examine our results for the chirp mass from expression \eqref{eq:chirpwdmd} even with a WDWD picture (see Sec.~\ref{sec:orbgws}).

\subsubsection{CHIME J0630+25} \label{sec:j0630}

A system where the possibilities described above are questionable is the shortest-period LPT, J0630 \cite[$P \approx 421$~s;][]{dong24}. 
This is for two main reasons, one being that the best-fit period derivative\footnote{{We remark, however, that only in cases where a timing glitch is excluded is a negative $\dot{P}$ supported; see Table 2 in \protect\cite{dong24}.}}, $\dot{P} = (-7.8 \pm 1.4) \times 10^{-13} \text{ ss}^{-1}$ at 1$\sigma$, would indicate a bound of $\mathcal{M} \lesssim 0.1 M_{\odot}$ if orbital evolution is orchestrated by GW losses \citep{peters64}:
\begin{equation} \label{eq:tension}
\begin{aligned}
\dot{P}_{\rm GW} &= -\frac{96 G^{5/3} (4 \pi^2)^{5/3}}{5 c^5} P_{\rm orb}^{-5/3} \mathcal{M}^{5/3}  \\
&\approx - 5 \times 10^{-11} \left( \frac{P_{\rm orb}}{421 \text{ s}} \right)^{-\tfrac{5}{3}} \left( \frac{\mathcal{M}}{0.5 M_{\odot}} \right)^{\tfrac{5}{3}} \text{ss}^{-1},
\end{aligned}
\end{equation}
{for Newton's constant $G$ and speed of light $c$.}
Secondly, the lack of an {obvious} optical counterpart is concerning owing to the closeness of the source ($d \sim 100$~pc).
{Notably, a counterpart should be fainter than 22 magnitudes based on nondetections in Pan-STARRS \citep{pans02}, and many WDWDs are brighter. 
For example, ZTF J153932.16+502738.8 (J1539) has an orbital period of $\sim 7$~min and is optically visible despite its $\sim 2.3$~kpc distance \citep{bur19}.} 
Dust obscuration or binary inclination could alleviate this tension to some 
{extent, as could 
additional optical monitoring given the relatively poor localization of the source,} 
{though these factors still challenge a WDWD interpretation.}
On the other hand, the fact that the best-fit $\dot{P}$ is \emph{negative} hints at a binary, as emphasized by \cite{dong24}. 
{Moreover, J1539 exhibits orbital decay that is consistent with equation \eqref{eq:tension}, providing evidence for the viability of a double-compact case \citep{bur19}.} 

All considered, we fix $\mathcal{M}_{\textrm{J}0630} = 0.1 M_{\odot}$ to avoid tension with expression \eqref{eq:tension}, remarking that orbital decay could be accelerated by tides or electromagnetic interactions (decreasing the permitted $\mathcal{M}$) or decelerated if dissipative torques transfer angular momentum from the primary to the orbit \cite[increasing it; see][]{marsh04}. We emphasize that this source, together with {A J1839 and} DA J1832 \cite[see][]{lee25,wang24}, is probably the least likely candidate for a binary LPT.
Nevertheless, 
{our key point} is that LISA can constrain binary scenarios in
{ways} electromagnetic data cannot: even if $\mathcal{M}$ was lower by an order of magnitude, the closeness of this source together with its short period would permit detection if $P = P_{\rm orb}$ (Sec.~\ref{sec:lisa}).

\subsubsection{GX J1627 and GPM J1839} \label{sec:gpm1839}

For the other shortest-period LPTs---GX J1627 ($P \approx 1091$~s) and GPM J1839 ($P \approx1318$~s)---the rough constraint implied by expression \eqref{eq:tension} is also important to consider {(see also Sec.~\ref{sec:j1634})}. {There are only loose period derivative estimates for the former object} \cite[$|\dot{P}| \lesssim 10^{-9} \text{ ss}^{-1}$;][]{hw22}, and 
{therefore it} poses no issue. 
The latter, by contrast, has a fairly tight and notably positive derivative quoted at $\dot{P} \lesssim 3.6 \times 10^{-13} \text{ ss}^{-1}$ \citep{hw23}. 
{However, inspection of extended-data figure 4 from} \cite{hw23} shows
that a \emph{negative} value of $\dot{P} \approx - 10^{-12} \text{ ss}^{-1}$ cannot be excluded by the timing analysis at 3$\sigma$ 
{nor} $\dot{P} \approx - 10^{-13} \text{ ss}^{-1}$ at 1$\sigma$. 
With this in mind, we restrict ourselves to light progenitors for this object by considering $\mathcal{M}_{\textrm{GPM J}1839} \leq 0.15 M_{\odot}$. This essentially precludes a LISA detection, unless $P \neq P_{\rm orb}$ or 
{the evolution is not dominated by GW losses, in which case expression \eqref{eq:tension} would not apply.}

GPM J1839 stands out amongst LPTs as this source has displayed a high duty cycle ($\sim 25\%$) going back to 1988 \citep{hw23}. In a neutron-star scenario, this points towards aggressive crustal activity despite being old and `cold' \citep{coop24}, which could be used to implicitly constrain magnetar evolution \cite[e.g.][]{deh20}. This is supported by observations of linear-to-circular polarization conversion taking place in a manner remarkably reminiscent of the transient magnetar XTE J1810--197 \citep{men25}. If instead confirmed as a binary, independently of whether $P = P_{\rm orb}$, the fact that the radio luminosity exceeds that of AR Sco by several orders of magnitude indicates that pulsations from binaries involving dwarfs could be highly variable, important to account for in future radio surveys. Dedicated X-ray monitoring would also
{help constrain the nature of the source, as the current upper limit of} $L^{\rm max}_{\rm X} \sim 10^{33}$~erg/s is relatively weak.

\subsection{Neutron stars plus one} \label{sec:nsp1}

Except where the above exclusions apply, the chirp mass could be considerably higher if invoking a compact neutron-star/white-dwarf binary like PSR J1141--6545 \cite[$P_{\rm orb} \approx 4.7$~hr and $\mathcal{M} \sim 1\, M_{\odot}$;][]{bailes03} or a neutron/helium star binary like PSR J1928+1815 \cite[$P_{\rm orb} \approx 3.6$~hr and $\mathcal{M} \sim 1.2 M_{\odot}$;][]{yang24}. 
Binaries involving mature neutron stars are particularly difficult to constrain with multiwavelength data, much like isolated objects. 
{The pulsation periods may} correspond 
{to the primary’s rotation rate} in such cases -- and indeed it has been argued that mass-loaded winds, magnetospheric twists, or interactions with a fallback disk could allow magnetars to reach ultra-long periods \citep{ben20,ben23,sdp25} -- {but a binary involving at least one neutron star is viable for many LPTs \cite[e.g.][]{lyman25}}. 
The final (most optimistic) scenario we consider corresponds to a J1141-like chirp mass of
\begin{equation} \label{eq:chirpns}
\mathcal{M}_{\rm NS} \approx 1 M_{\odot},
\end{equation}
where the subscript indicates {multiple possibilities for a compact companion (helium dwarf, white dwarf, or neutron star)}. 
For example, \cite{tur05} suggested that GCRT J1745--3009 could contain a double neutron-star system with $\mathcal{M} \approx 1.13 M_{\odot}$ and $P = P_{\rm orb}$ where radio emission originates from shocks within the interaction zone of the primary's wind and the companion's magnetosphere.

\subsubsection{CHIME/ILT J1634+44} \label{sec:j1634}

{This source 
exhibits the unusual property that its emission is almost 100\% circularly polarized
\citep{dong25}, 
although some pulses are nearly completely linearly polarized \citep{bloot25}.
This contrasts with other LPTs and 
further emphasizes the likelihood of (at least) two distinct progenitor categories.
An important feature of J1634 
is the confident detection by CHIME of a \emph{negative} period derivative,
$\dot{P} \approx -9.03(11) \times10^{-12} \text{ ss}^{-1}$ \citep{dong25}. 
Despite the absence of any multiwavelength counterparts (with an X-ray limit of $L^{\rm max}_{\rm X} \lesssim 10^{32}$~erg/s), this provides strong evidence for a binary.}

{While the observed pulsation period $P = 841.25$~s may correspond to the orbit, the existence of occasional, interpulse-like peaks with period $\sim 4206$~s could be related to a spin-orbit resonance \citep{bloot25}. 
In particular, these secondary `bursts' were observed only at certain phases with different Stokes offsets depending on the observational epoch. 
The modulation period of $P = 2103.1$~s was interpreted by \cite{bloot25} to be the orbital period assuming a 5:2 spin-orbit resonance.} 

{Since both \cite{bloot25} and \cite{dong25} mention that the radio data alone cannot strictly exclude an orbital period of $P = 841.25$~s or a larger modulation period, we consider} two scenarios for this source, with either $P_{\rm orb} = 841.2$~s or $P_{\rm orb} = 2103.1$~s. 
The period derivative quoted above then translates into chirp masses of $\mathcal{M} \approx 0.36 M_{\odot}$ or $\mathcal{M} \approx 0.91 M_{\odot}$ through expression \eqref{eq:tension}, respectively, assuming GW emissions dominate angular momentum losses. 
The former 
aligns better with a WDMD or WDWD, while the latter is more consistent with a neutron star `companion'. 
{We consider multiple scenarios for the $P = 2103.1$~s case, since a unipolar inductor or tides (for instance) could accelerate inspiral and thus reduce the inferred $\mathcal{M}$ (see also Sec.~\ref{sec:future}).}

We close this section by stressing that we have not attempted to present a complete survey of LPT scenarios, as these can be found in the respective discovery papers and literature \cite[see, e.g.,][]{rea22,cz24}. 
Indeed, our goal here concerns GW visibility, for which the exact binary makeup is unimportant (though is, of course, important 
{for the \emph{implications} of detections)}
Even if certain types of binarity may be disfavored in individual cases, we consider a range of chirp masses to estimate S/Ns. Larger or smaller $\mathcal{M}$ values can be easily substituted using the estimates provided in the following sections.

\section{Orbital gravitational waves} \label{sec:orbgws}

\begin{figure*}
\centering
  \includegraphics[width=\textwidth]{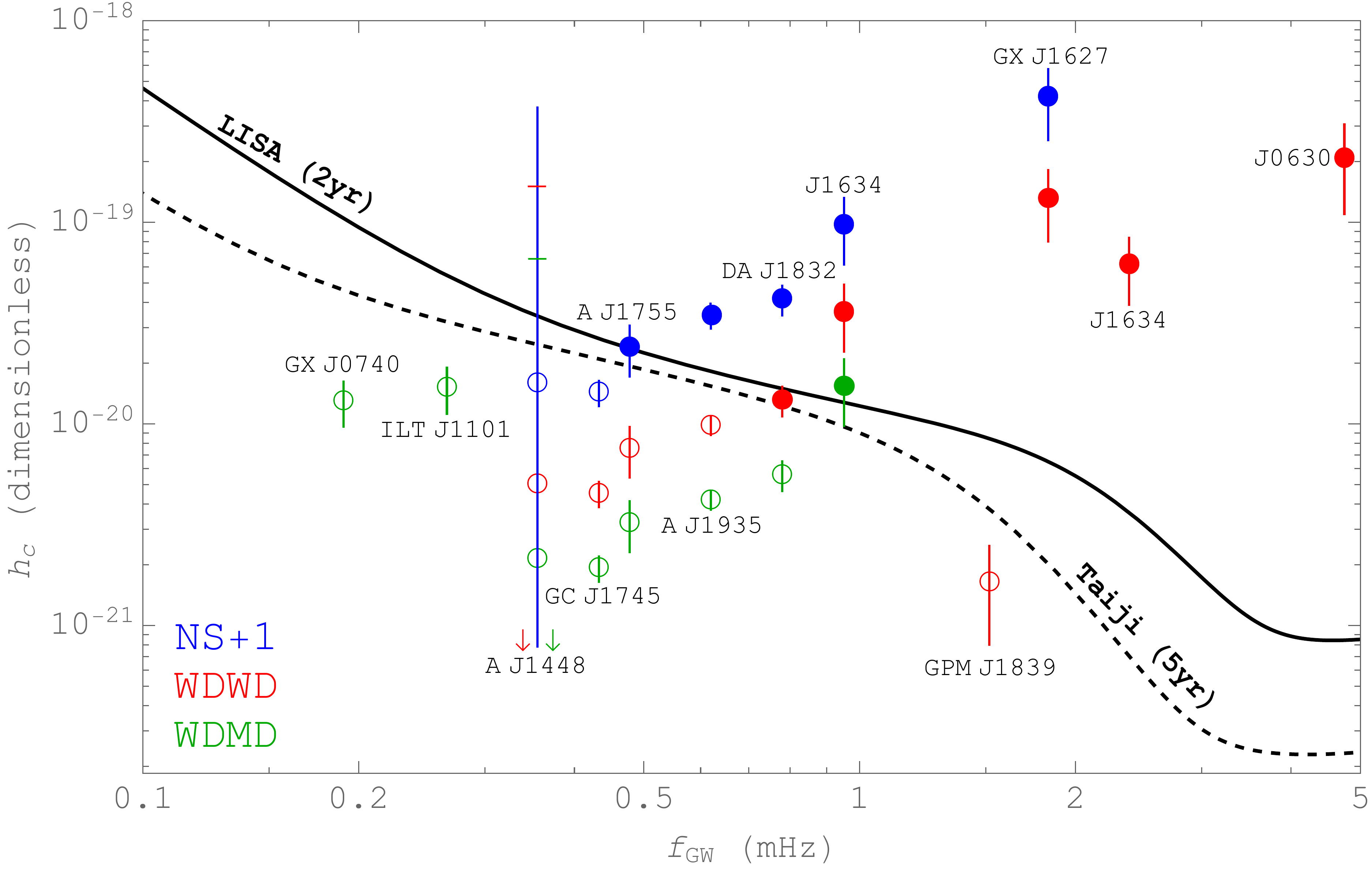}
  \caption{{Characteristic strains, equation~\eqref{eq:lisahc}, for LPTs listed in Tab.~\ref{tab:orbdata} (see legends) assuming $P = \Po$ and $T = 2$~yr. 
  Objects are color-coded according to different scenarios corresponding to chirp masses $\mathcal{M}$ set by expressions \eqref{eq:chirpwdmd} (green), \eqref{eq:chirpwdwd} (red), or \eqref{eq:chirpns} (blue). 
  Exceptions are CHIME J0630 (Sec.~\ref{sec:j0630}), GPM J1839 (Sec.~\ref{sec:gpm1839}), and J1634 (Sec.~\ref{sec:j1634}), where orbital braking and Roche-lobe maxima restrict the component masses and companion makeup, respectively, if there is no active mass transfer. 
  We consider two orbital periods for J1634, as detailed in Sec.~\ref{sec:j1634}. 
  Filled symbols correspond to cases which exceed or overlap with the LISA or Taiji strain curves, taken from data provided by \protect\cite{rob19} and \protect\cite{luo20}, respectively. 
  The noise curve for Taiji is shown instead for the mission lifetime of 5 years to avoid visual clutter. 
  Error bars on A J1448 are large due to the lack of distance constraints \protect\citep{ak25}.
  }}
  \label{fig:LISALPTs}
\end{figure*}

The intrinsic GW amplitude of a circular binary reads 
\begin{equation} \label{eq:lisa0}
h_{0} \approx 2 \times 10^{-22} \left( \frac{1 \text{ hr}} {\Po} \right)^{2/3} \left( \frac {\mathcal{M}} {0.5 M_{\odot}} \right)^{5/3} \left( \frac {1 \text{ kpc}} {d} \right).
\end{equation}
Assuming that the pulse period is that of the orbit, we use the data collated in Table~\ref{tab:orbdata} to estimate $h_{0}$ from equation \eqref{eq:lisa0} for each LPT using the chirp mass spread considered in Sec.~\ref{sec:multi}. Notably, the period derivative is moderate in most systems \cite[$|\dot{P}|_{\rm max} \lesssim 10^{-9} \text{ ss}^{-1}$;][]{cz24}, meaning that GW signals should be essentially monochromatic over multi-year observational windows, $T$ {(i.e., the folding time)}. This allows for an accumulation of signal power over many orbital cycles, $N = T f_{\rm GW} = 2T/P_{\rm orb}$. Averaging over orbital orientations and polarizations, the characteristic strain $h_{c}$ felt by a detector can thus be estimated through $h_{c} \approx \sqrt{2 N} h_{0}$ \citep{finn00}, with value
\begin{equation} \label{eq:lisahc}
\begin{aligned}
h_{c} \approx&\,\, 2.1 \times 10^{-20} \left( \frac{1 \text{ hr}} {\Po} \right)^{7/6} \left( \frac {\mathcal{M}} {0.5 M_{\odot}} \right)^{5/3}\\
&\qquad \times \left( \frac {T}{0.5 \text{ yr}} \right)^{1/2} \left( \frac {1 \text{ kpc}} {d} \right).
\end{aligned}
\end{equation}

Using the LISA design specifications from \cite{lisa17}, we adopt the noise power spectral density ($S_n$) fits from \cite{rob19}, which include intrinsic noise (from single-link optical metrology and test mass acceleration) as well as Galactic confusion noise from unresolved sources, to estimate LPT detectability. 
{For Taiji, the same Galactic confusion noise profile is used, but with the design specifications 
provided by \cite{luo20} and \cite{cl25}.}
Figure~\ref{fig:LISALPTs} shows characteristic strains $h_{c}$,
{calculated using} equation \eqref{eq:lisahc},
as functions of GW frequency for 
{the} objects listed in Tab.~\ref{tab:orbdata}, assuming an observation time of two years.
We find that {three} of the known LPTs could be 
detectable by LISA {and Taiji} (filled symbols),
with {several} others being marginal under favorable conditions (see Sec.~\ref{sec:lisa}). 
{It is noteworthy that} the most promising candidates are also among the most intriguing. 

\textit{CHIME J0630.}~Given its proximity to Earth, insights on local supernovae rates could be gleaned if this source is a neutron star, a position which would be firm if no detection is made shortly after LISA launch. The period would thus correspond to rotation, requiring that some magnetars spindown faster than stipulated by pure dipole braking \cite[e.g. via mass-loaded winds;][]{ben20}. 
An excess of nearby neutron stars (and magnetars in particular) could also call for revisions to population synthesis models \cite[though see][]{pop03}. 
While a nearby white dwarf is not particularly unusual,
\cite{tauris18} estimates that for S/Ns $\gtrsim 100$, $\mathcal{M}$ could be measured to within $\lesssim 1\%$ from which constraints on binary makeup, 
and possibly even pre-Newtonian effects predicted by some modified theories of gravity \citep{ly19}, follow.

{\textit{ILT/CHIME J1634.}~Given both its unusual polarization properties and confidently negative $\dot{P}$, this source is a compelling candidate for a binary. 
The orbital frequency is somewhat uncertain,  
and multiple scenarios with double-degenerates are viable \citep{dong25,bloot25}.
A joint LISA/Taiji detection would narrow down the possibilities by not only measuring the chirp mass but $f_{\rm GW}$ directly. 
Given that microstructure was observed within the pulses resembling that from XTE J1810--197 and the long-period pulsar PSR J0901--4046 \citep{caleb22,dong25}, it is possible that this system consists of a tight binary with a magnetar. 
A detection may thus shed light on binary interactions in systems with strong magnetic fields \cite[see, e.g.,][]{gs21}. 
The system should also be detectable [$\text{S/N}(4\text{ yr)} \geq 10$] if $P_{\rm orb}$ corresponds instead to the longest `burst-like' emissions ($P \sim 4036$~s) for chirp masses $\mathcal{M} \gtrsim 1 M_{\odot}$.}

\textit{GX J1627.}~Discovered in 2018 archival data from the Murchison Widefield Array, this object has since
{remained} silent \citep{hw22}.
If interpreted as a neutron star, because the spindown luminosity, $L_{\rm sd} \lesssim 10^{28}$~erg/s, is much lower than the radio luminosity, $L^{1.4 \text{GHz}}_{\nu} \sim 4 \times 10^{31}$~erg/s, it could not be rotation powered. 
If magnetically powered, the deep X-ray limits set by Chandra \cite[$L^{\rm max}_{\rm X} \sim 10^{30}$~erg/s;][]{rea22}
{imply} that a non-detection by LISA could indicate that (some) LPTs populate a magnetar branch
distinct from soft gamma repeaters and anomalous X-ray pulsars \citep{sdp25}. 
For a dwarf, the lack of radio reactivation 
{suggests either a choked magnetosphere (e.g. due to accretion) or inconsistently beamed emissions, either of which could provide insight into tight-binary interactions.}

\textit{DA J1832.} Because this LPT has shown contemporaneous X-ray bursting with radio pulsing \citep{wang24}, a magnetar progenitor is favored over a binary \citep{sdp25}. If confirmed as a binary however, sporadic accretion could be responsible for igniting a burst of comparable luminosity to that seen from DA J1832 in February 2024 ($L_{\rm X} \sim 10^{33}$~erg/s), though it would be difficult to explain the lack of any quiescent emissions \cite[see][]{muk17}. If the object enters into an outburst phase in future, searches for absorption lines could be useful to constrain its nature. The fact that the source was radio-loud for a $\sim$~year is also uncharacteristic for transitional pulsars (for instance), and so a detection could insist on revisions to theoretical models from a variety of directions.

\textit{A J1935.} This source displays a unique property amongst LPTs, being that of state switching: there are pulse states with high-degrees of linear polarization, others which are predominantly circular, and a null state \citep{caleb24}. 
A high degree of linear polarization and frequent nulling is rather reminiscent of radio-loud magnetars. 
For an isolated object, a highly-dynamic magnetosphere would be required to explain this temperamental behavior; the evolution could be driven, for example, by helicity injections from the plastic motions of crustal platelets \cite[e.g.][]{belo09}.

\subsection{Signal-to-noise ratios} \label{sec:lisa}

\begin{table}
\centering
\caption{S/Ns for LISA for the systems listed in Tab.~\ref{tab:orbdata} for $T = 0.5$~yr (middle column) or $T = 4$~yr (right) in either the least (largest distances and smallest chirp masses) or most (bracketed values) optimistic scenarios.}
\hspace{-1.2cm}\begin{tabular}{lcc}
 \hline
 \hline
Source name & S/N (0.5~yr; max) & S/N (4~yr; max) \\
\hline
CHIME J0630+25 & 84.0 (233) & 238 (660)  \\
J1634+44 & 1.37 (43.4) & 3.86 (123)  \\
GLEAM-X J1627 & 31.0 (222) & 87.8 (627) \\
GPM J1839--10 & 0.27 (0.58) & 0.77 (1.65) \\
DA J1832 & 0.42 (4.33) & 1.18 (12.3) \\
ASKAP J1935 & 0.22 (2.03) & 0.63 (5.73) \\
ASKAP J1755 & 0.09 (1.20) & 0.26 (3.40) \\
GCRT J1745--3009 & 0.05 (0.50) & 0.14 (1.41) \\
ASKAP J1448 & 0.05 (7.97) & 0.15 (22.5) \\
ILT J1101+5521 & 0.13 (0.22) & 0.37 (0.62) \\
GLEAM-X J0704 & 0.05 (0.09) & 0.15 (0.25) \\
\hline
\hline
\end{tabular} 
\label{tab:S/Ndata}
\end{table}

In general, a confident detection requires an S/N of at least $\sim$~10. While detection odds improve for sources with known sky locations, we provide conservative estimates by avoiding complications related to barycentric and other corrections.
Borrowing formula (26) from \cite{rob19}, the orientation- and polarization-averaged S/N is estimable through
 \begin{equation} \label{eq:S/N}
\text{S/N} \approx \frac {8 \pi^{2/3} G^{5/3} \mathcal{M}^{5/3} \fGW^{2/3}} {\sqrt{5} c^4 d} \sqrt{ \frac {T} {S_{n}(\fGW)}}\ .
\end{equation}
We calculate expression \eqref{eq:S/N} for sources listed in Tab.~\ref{tab:orbdata}. The results are shown in Table~\ref{tab:S/Ndata} 
for two scenarios: 
(i) using the maximum inferred distances (through dispersion measures) and lowest values of $\mathcal{M}$ considered in Sec.~\ref{sec:multi}, and 
(ii) an optimistic case where the masses and distances take their upper and lower confidence bounds, respectively. 
The latter scenario is indicated by bracketed values in the second ($T=0.5$~yr) and third ($T=4$~yr) columns of Tab.~\ref{tab:S/Ndata}.

As expected visually from Fig.~\ref{fig:LISALPTs}, taking $T \lesssim 0.5$~yr practically guarantees that CHIME J0630 and GX J1627 would be identifiable in the LISA data stream if $P = P_{\rm orb}$. 
While hopes for other LPTs are low, there is some call for optimism for DA J1832 and A J1935 if the systems contain heavier stars, allowing them to reach significant S/Ns when accounting for their sky locations. 
For instance, if A J1935 contains a neutron star and a reasonably heavy companion, its well-constrained sky location could allow for a boost to the S/N by a factor $\sim 2$, which would give a final value of $\text{S/N}(4 \text{yr}) \sim 12$: sufficient to claim a confident detection. 

\subsection{Improvements via a network of detectors} \label{sec:network}

Combining data from LISA with Taiji {(and/or TianQin)} could also lead to non-negligible increases in the S/N for each source, perhaps enabling detections in marginal scenarios like a WDWD for DA J1832. 
{If we assume that the response of a network to some impinging GW is phase coherent in each detector, we can estimate the joint S/N through \citep{finn01}}
\begin{equation} \label{eq:snformula}
    \text{S/N} \approx \sqrt{(\text{S/N})_{\rm LISA}^2 + (\text{S/N})_{\rm Taiji}^2 + (\text{S/N})_{\rm TianQin}^2}.
\end{equation}
{As discussed by \cite{cl25}, given the similarities between the Taiji and LISA noise curves (see Fig.~\ref{fig:LISALPTs}) we expect the individual $\text{S/N}$ values to be comparable for a given source. 
The S/N measurable by TianQin will be lower for an LPT. 
The total S/N \eqref{eq:snformula}, for any given LPT, may thus increase by a factor $\gtrsim \sqrt{2}$ relative to the results quoted in Tab.~\ref{tab:S/Ndata} (see also Sec.~\ref{sec:future}).}

{If the location of a given source is pinpointed in future, further enhancements may be expected since (correlated) confusion noises can be reduced. 
We anticipate a factor $\sim 2$ improvement in this case \citep{schutz11,rob19}. 
Although a thorough statistical analysis lies beyond the scope of this article, such an increase could bolster the S/N past 10 in 
several marginal cases 
(e.g., a binary neutron-star scenario for A J1935 or a WDWD scenario for A J1448), thereby enabling a confident detection.}

{Aside from the obvious benefits anticipated from equation \eqref{eq:snformula}, there are indirect benefits that a network would offer. 
Given that LISA and Taiji will be positioned with large separations in space while in heliocentric orbits, combining their data streams could allow for a source to be accurately localized over the course of a long observational campaign \cite[in terms of both distance and solid angle;][]{ruan20}. 
This could be especially useful to reduce error boxes to help in identifying optical and X-ray counterparts. 
A network of detectors can also veto glitch events \citep{schutz11}. 
While detector glitches are less important for long-lived signals, a range of scenarios could disable a detector for a period of time. 
The impact of offline time for a given detector would be mitigated with a network.}

\section{Outlook for future surveys} \label{sec:future}

{We have thus-far considered the detection prospects of known LPTs. 
Given the rapidity with which the field is developing, additional sources are likely to be discovered in future.}

\begin{figure}
\centering
  \includegraphics[width=0.49\textwidth]{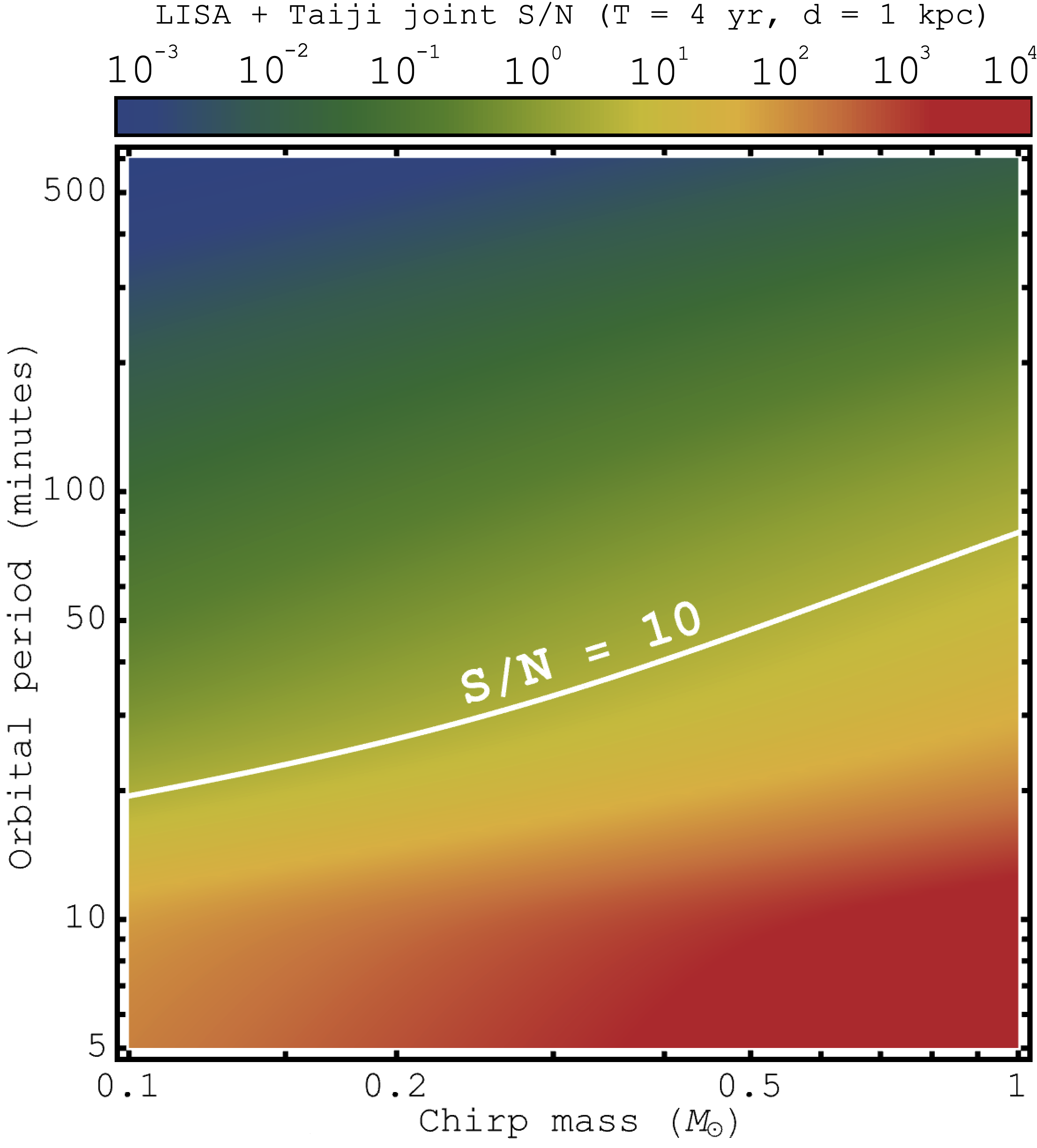}
  \caption{{Sky-averaged signal-to-noise ratio (color scale), as a function of chirp mass and orbital period, for hypothetical LPTs jointly-observed by LISA and Taiji. An observation time of $T = 4$~yr and a distance of $d=1$~kpc are assumed. Redder shades indicate higher S/N values.
  }}
  \label{fig:generalsnr}
\end{figure}

{As is clear from Tab.~\ref{tab:orbdata}, there is no obvious clustering of periods in LPTs at present and the known range spans over an order of magnitude. 
In fact, a candidate LPT with a period of $P \sim 3.7$~days,
containing a K7V dwarf, was recently discovered \cite[J180526--292953;][]{frail25}. 
Nevertheless, we can investigate LPT detectability in general terms.
Figure~\ref{fig:generalsnr} shows sky-averaged S/Ns for a joint observation spanning 4~years for sources located at a distance of 1~kpc. 
In the bottom-right corner, corresponding to large $\mathcal{M}$ and short $P_{\rm orb}$, the highest S/N values are obtained ($\sim 10^{4}$). 
Such cases may apply to systems like GX J1627 in optimistic scenarios involving a neutron star (see Tab.~\ref{tab:S/Ndata}), with leftward shifts for a WDWD or WDMD. 
The overlaid white curve marks $\text{S/N} = 10$: LPTs below this line are likely detectable \citep{rob19}. 
For example, for a WDWD with $\mathcal{M} = 0.5 M_{\odot}$ from \eqref{eq:chirpwdwd}, orbital periods of $P_{\rm orb} \lesssim 41$~min (similar to that of DA J1832) would be needed for LISA+Taiji to detect a signal. 
If the source were located instead at a J0630-like distance of $d = 100$~pc, periods reaching $P_{\rm orb} \approx 114$~min would still enable $\text{S/N} > 10$.
If a $P$--$\mathcal{M}$ correlation reveals itself in future, it may be possible to hunt for LPTs even without pulse data. 
In particular, candidates could be identified via GWs and localized with a network \citep{ruan20}, which could be used to guide radio surveys.}

{The leading explanation for pulsations in binary LPTs involves a relativistic cyclotron maser \citep{qu25}. 
This mechanism requires the formation of a unipolar inductor, which will also work to drain orbital angular momentum together with GW losses \eqref{eq:tension} \cite[e.g.][]{lai12}. 
The energy dissipation rate for GWs,}
\begin{equation}
    \dot{E}_{\rm GW} \approx 10^{32} \left(\frac{\mathcal{M}}{0.4 M_{\odot}}\right)^{10/3} \left(\frac{60 \text{ min}}{P_{\rm orb}}\right)^{10/3} \text{erg/s},
 \end{equation}
{is typically large with respect to observed radio luminosities \cite[e.g. $L_{\nu} \sim 10^{31}$~erg/s for the brightest single peaks in GX J1627;][]{hw22}. 
Since perfectly-efficient conversion from electromagnetic dissipation to radio emission is unlikely though, some or even most of the orbital decay could be attributed to an inductor \citep{willes04}. 
A detection of $\dot{P}$, together with $\mathcal{M}$, could be used to assess this and implicitly constrain the radio emission mechanism. 
If the $\dot{P}$ and $\mathcal{M}$ values match closely with equation \eqref{eq:tension}, pulsations would have to instead be fuelled by one of the individual objects. 
This could provide a way to identify magnetars in binaries, none of which are known to date \cite[though see][for a discussion on LS I+61{\textdegree}303]{sg22}. 
By contrast, active mass transfer can widen the orbit and counterbalance GW losses. 
Future data could thus also be used to implicitly constrain the mass accretion rate to see if/how LPTs fit within the population of radio-loud catacyslmic variables \citep{rid23}.}

\section{Conclusions} \label{sec:discussion}

LPTs {are} mysterious, with some studies favoring isolated neutron stars and others favoring WDMDs or compact binaries \cite[see, e.g.,][]{rea22,cz24}. 
We have argued here that \emph{if} the pulsation period in LPTs matches,
{or is close to, the orbital period, then}
at least a few should be visible to LISA and its sister spacecraft (Tab.~\ref{tab:S/Ndata}). 
If electromagnetic data remain inconclusive, GW observations could help unveil their nature. 
Importantly, among the known LPTs, those that may be brightest in the LISA band correspond to objects with uncertain classification (Fig.~\ref{fig:LISALPTs}). 

{Although we have presented a number of progenitor scenarios, the value of $\mathcal{M}$ is unknown in binary LPTs and must be searched over in a LISA/Taiji survey. 
This also applies to the distance
\cite[see][]{price21}, meaning it is the combination $\mathcal{M}^{5/3} d^{-1}$ featuring in expression \eqref{eq:S/N} that is particularly uncertain. 
Given the precision with which the period can/has been measured due to the timing of pulses though, it is unlikely that another source could be confused for a given LPT in the respective data streams upon Fourier binning (as the bin size $\Delta f \sim 1/T \ll f_{\rm GW}$).
This applies especially to coherent searches, which are best-suited to reject coincident background signals \citep{klim05}. 
As highlighted at the beginning of Sec.~\ref{sec:orbgws}, the period derivative is gradual in LPTs. 
This means that the signal should be essentially monochromatic and it is thus straightforward to construct many phase templates to carry out matched filtering \citep{schutz11}. 
In general, the uncertainty in $\dot{f}_{\rm GW}$ can be estimated as $\Delta \dot{f}_{\rm GW} \approx 4.3/(T^2 \times \text{S/N})$ \citep{tak02}, from which the chirp mass uncertainty can be deduced as $\Delta \mathcal{M} / \mathcal{M} \approx 3/5 \times \Delta \dot{f}_{\rm GW}/ \dot{f}_{\rm GW}$ \citep{lau20}. 
We refer the interested reader to the above references for discussions on parameter surveys and search strategies.}

While some implications of (non-)detections {have been highlighted}, a wealth of physical processes may operate in LPTs.
For instance, linking radio luminosities to electromotive losses implies a minimum strength for the primary's magnetic field. Assuming a maser-like emission mechanism, some systems appear 
to require
GigaGauss strengths \citep{qu25}. 
While dynamo action may amplify an internal field \citep{sch21}, the magnetospheric field may not reflect the interior strength because magnetic burial -- a process where field lines are equatorially advected by infalling plasma -- will reduce the global dipole moment during accretion while creating strong multipoles \citep{suvm20}. 
Typical accretion rates in tight-binary polars are $\sim 10^{-10} M_{\odot} \text{ yr}^{-1}$ \citep{pala20} implying that, for a C/O dwarf with $M_{\star} \sim 0.8 M_{\odot}$, 
the accretion timescale matches that of Ohmic diffusion ($\tau_{\Omega}$) at depths corresponding to densities of $\sim 10^{5} \text{ g cm}^{-3}$ where \cite{cumm02} estimates $\tau_{\Omega} \lesssim 1$~Gyr. 
This suggests a timeline for radio activation post detachment if the interaction zone lies at a polar latitude, with the reverse applying near the equator if magnetic flux is compressed there. 
Resistive relaxation simulations would 
help elucidate such interactions and their connections with diamagnetic screening, crystallization, and convection.

\section*{Acknowledgements}
{AGS thanks Adam Dong for correspondence on the nature of CHIME J0630 and ILT/CHIME J1634. We thank the anonymous referee for their helpful comments.} 
Support provided by the Conselleria d'Educaci{\'o}, Cultura, Universitats i Ocupaci{\'o} de la Generalitat Valenciana through Prometeo Project CIPROM/2022/13 is gratefully acknowledged.
CD acknowledges the Ministerio de Ciencia, Innovación y Universidades, co-funded by the Agencia Estatal de Investigación, the Unión Europea (FSE+), and the Universidad de Alicante, through fellowship JDC2023-052227-I funded by MCIU/AEI/10.13039/501100011033 and the FSE+.
CD and JAP acknowledge support from the 
grants PID2021-127495NB-I00 and ASFAE/2022/026 funded by MCIN AEI/10.13039/501100011033 and by the European Union, via NextGenerationEU (PRTR-C17.I1).

\label{lastpage}

\end{document}